\documentclass[prb,showpacs,superscriptaddress,twocolumn,notitlepage]{revtex4-1}

\usepackage{graphicx, subfigure}

\usepackage{amsmath}
\usepackage{amsfonts}
\usepackage{graphicx}

\usepackage{tabularx}

\usepackage{natbib}
\usepackage[final=true]{hyperref}

\usepackage{color}
\usepackage[usenames,dvipsnames]{xcolor}

\usepackage{multirow}

\newcommand{\ml}{(\lambda)}
\newcommand{\sca}{_{\mathrm{sca}}}

\newcommand{\fs}{_{\mathrm{Diff}}}
\newcommand{\s}{_{\mathrm{Sphere}}}

\newcommand{\scau}{^{(\mathrm{sca})}}
\newcommand{\absu}{^{(\mathrm{abs})}}

\newcommand{\AbsW}{\mathrm{Abs}_{\mathrm{IS}}}

\newcommand{\AbsWO}{\mathrm{Abs}_{\mathrm{N}}}
\newcommand{\Abs}{\mathrm{Abs}}

\newcommand{\WO}{_{\mathrm{Par}}}

\begin{document}
\title{Evidence of absorption dominating over scattering in light attenuation by nanodiamonds}

\author{S.~V.~Koniakhin}
\email{kon@mail.ioffe.ru}
\affiliation{Institut Pascal, PHOTON-N2, University Clermont Auvergne, CNRS, 4 avenue Blaise Pascal, 63178 Aubi\`{e}re Cedex, France}
\affiliation{Ioffe Institute, 194021 St.~Petersburg, Russia}
\affiliation{St. Petersburg Academic University - Nanotechnology Research and Education Centre of the Russian Academy of Sciences, 194021 St. Petersburg, Russia}

\author{M.~K.~Rabchinskii}
\affiliation{Ioffe Institute, 194021 St.~Petersburg, Russia}

\author{N.~A.~Besedina}
\affiliation{Ioffe Institute, 194021 St.~Petersburg, Russia}
\affiliation{St. Petersburg Academic University - Nanotechnology Research and Education Centre of the Russian Academy of Sciences, 194021 St. Petersburg, Russia}

\author{L.~V.~Sharonova}
\affiliation{Ioffe Institute, 194021 St.~Petersburg, Russia}

\author{A.~V.~Shvidchenko}
\affiliation{Ioffe Institute, 194021 St.~Petersburg, Russia}

\author{E.~D.~Eidelman}
\affiliation{Ioffe Institute, 194021 St.~Petersburg, Russia}
\affiliation{St. Petersburg Chemical Pharmaceutical Academy, 197022 St. Petersburg, Russia}

\begin{abstract}

We show an experimental evidence of the domination of absorption over scattering in absorbance spectra of detonation nanodiamonds. We perform the absorbance measurements on the UV-Vis spectrophotometer equipped with integrating sphere and compare them with conventional absorbance spectra. Additionally, we measure the scattering light intensity at the cuvette side wall (scattering at 90 degrees angle). The obtained experimental data were interpreted using the simulations of photon random walk in turbid media and Kubelka-Munk approach. The scattering cross sections and indicatrices were obtained by Mie theory.

We discover that despite being very close to $\lambda^{-4}$ power law (like Rayleigh scattering) the light extinction by the primary 4 nm diamond crystallites is due to absorption only and scattering can be neglected. That is the reason why previously absorption and scattering contributions were confused. The scattering is governed only by the agglomerates of 100 nm and larger in size remaining in the hydrosols and their fraction can be effectively controlled by centrifugation. Only Mie theory reproduces correctly the close to $\lambda^{-2}$ scattering by the agglomerates accounting for the weird interplay between their size, fractal dimension, and dielectric properties. Finally, using the obtained absorbance spectra we estimate the fraction of non diamond phase in nanodiamonds and their agglomerates.

\end{abstract}

\maketitle
\section{Introduction}

Nanodiamonds are one of the most unique nanoparticles being currently investigated due to their exceptional mechanical, heatь and optical properties inherited from the bulk diamond\cite{hui2010nanodiamonds,ho2010applications, schrand2009nanodiamond,mochalin2012properties,arnault2017nanodiamonds}. Nanodiamonds exhibit high thermal conductivity \cite{kidalov2007thermal} and mechanical strength, can contain  bright, long-lived and controllable color centers\cite{boudou2009high,Chen2017,stacey2012depletion,fujiwara2018tracking,tisler2009fluorescence}. Current and future applications of nanodiamonds include NV defects-based quantum computing \cite{neumann2008multipartite,robledo2011high,bernien2012two}, composite materials creation \cite{behler2009nanodiamond,maitra2009mechanical,chen2018mechanical,kidalov2007thermal,kurkin2016polymer,guillevic2019nanodiamond}, sustainable energy\cite{tafti2016novel,hejazi2018using},  bioimaging\cite{faklaris2009photoluminescent,nunn2018fluorescent}, and drug delivery\cite{zhu2017nanodiamond}. Along with manufacturing, modifying, and investigating the high-pressure high-temperature \cite{boudou2013fluorescent,stehlik2015size,nunn2018fluorescent}, bead milling\cite{boudou2009high}, laser synthesis \cite{baidakova2013structure}, and even extraterrestrial\cite{shiryaev2011spectroscopic} nanodiamonds, among the most promising are the detonation nanodiamonds (DND)\cite{mochalin2012properties,shenderova2014detonation}.  Besides the powders, the most  important form for DND and other nanodiamonds are the water suspensions (hydrosols), easy-to-handle and native for chemistry and biology. Despite the serious progress that was achieved in the nanodiamond size control and fractionation\cite{stehlik2015size,stehlik2016high,dideikin2017rehybridization}, the hydrosols contain both individual primary crystallites and their agglomerates. Moreover, individual DND particles tend to form chains in hydrosols \cite{kuznetsov2018effect}.

To better understand the size distribution, structure, and phase composition of nanodiamonds, the optical experiments including measuring Absorbance (Abs) spectra are widely used \cite{tomita2002optical,vul2011absorption,aleksenskii2012optical,usoltseva2018absorption,volkov2012quantification,klemeshev2016static}. One of the most conventional and widely applied methods of nanodiamond characterization is dynamic light scattering (DLS) \cite{osawaDLS,osawa2008monodisperse,koniakhin2015molecular,aleksenskii2012applicability}, which requires precise knowledge of the optical parameters of the materials. This makes the deeper understanding of nanodiamond hydrosols optical properties highly desired. The surface effects closely connected with optical absorption are important for the manifestation of NV defects and for the quenching of their luminescence \cite{hauf2011chemical,ofori2012spin,loretz2014nanoscale}. 

Abs spectra (also referred as UV-vis spectra or Optical density spectra) of detonation nanodiamond hydrosols can be described as a superposition of light scattering and absorption \cite{vul2011absorption,aleksenskii2012optical}. The peculiar shape of the optical density spectra is thought to be defined by the interplay between these two effects, with the domination of scattering. For calculating the scattering cross section, the Rayleigh and Mie theories are used. Calculating the absorption cross section requires a model where nanoparticle electric polarization contains the imaginary part. Usually, the nanoparticle core-shell models with the presence of diamond-like core and graphite-like phase on the surface \cite{lucas1994multishell,tomita2002optical,vul2011absorption} are used, where the dielectric constant of graphite-like phase contains significant imaginary part giving the absorption. Numerical values of graphite-like dielectric constant are approximated with data for bulk graphite\cite{draine1984optical}. The calculations predict that the scattering dominates the absorption by a factor of 10 at shorter wavelengths (lower than 500 nm), whereas in red and near-IR regions the contribution of scattering and absorption becomes comparable \cite{vul2011absorption}.

Another approach for the determination of nanodiamond optical properties is based on ab-initio calculation of the nanodiamond electronic structure and derivation of the corresponding light absorption \cite{vantarakis2009interfacial}. These calculations allow accounting for such effects as surface reconstruction, presence of amorphous phase and carbon atoms with intermediate between sp$^2$ and sp$^3$ hybridization.

Here, we present an experimental evidence that 4 nm nanodiamonds dominantly \emph{absorb} light in all visible range, including near-UV and near-IR, which differs from the previous models suggesting that Abs spectra of DND hydrosols are mainly governed by the scattering. This picture generally remains valid even for agglomerates, where the absorption is determined to be comparable with the scattering. These results lead to a global change of the paradigm of the DND hydrosols Abs spectra interpretation.

The paper is organized as follows. In section \ref{ssec_samples}, we describe the preparation of the samples. The main quantity about which the present paper is composed is light intensity $I\sca$, scattered by nanodiamonds in the hydrosol and thus gone away from the cuvette. At a qualitative level, it is obvious that $I\sca$ positively correlates with the scattering cross section of the nanoparticles in hydrosol and negatively correlates with the absorption cross section. $I\sca$ can be addressed in three ways:

\begin{itemize}
\item Calculated via the difference between the Abs spectra measured with integrating sphere (IS) and without it (section \ref{ssec-uvvis}).
\item Detected straightly as the light intensity scattered at 90 degree angle through the cuvette side wall (section \ref{ssec_90deg}).
\item Via the numerical simulation of the photon random walks in the medium where scattering and absorption takes place (section \ref{sec_simulation}).
\end{itemize}

In section \ref{sec_results} we describe the obtained experimental and theoretical results and in section \ref{sec_discussion} we discuss them and establish a relation between all three approaches mentioned above. We will show that from the experimental results one can quantitatively conclude on the absorption and scattering contributions. The comparison of the experimental data and results of the simulation provides the quantitative level of extracting contributions to DND absorbance from scattering and absorption.




\section{Experiment}

\subsection{Samples}

\label{ssec_samples}

Preparation and physical-chemical characterization of the samples is described in details in Ref. \cite{dideikin2017rehybridization}. In brief, the preparation and purification protocol reads as follows. As an initial DND, the powder of an industrial DND was taken and an additional purification with the mixture of HF and HBr from inorganic impurities was done to obtain Z0 sample. DND Z0 powder was annealed in hydrogen at 600$^{\rm o}$ C for 3 hours for producing DND Z+ sample. DND Z- was obtained after annealing Z0 powder in air at 450$^{\rm o}$ C for 6 hours. Thus, three DND powders (Z0, Z+, and Z-) were obtained their names of the samples originate from the results of their Zeta-potential measurements \cite{dideikin2017rehybridization}. The difference in deagglomeration procedure was a reason of a difference in surface chemistry of the samples. According to the reported in Ref. \cite{dideikin2017rehybridization} XPS, XAS and FTIR data, DND Z- and Z+ are both grafted with CH and -COOH/-C(O)O- groups, however in different relations. DND Z- surface contains mainly carboxyls and lactones \cite{schmidlin2012identification}, while Z+ is hydrogenated. This picture correlates with the values of Zeta-potential. The important feature of the studied samples \cite{dideikin2017rehybridization}, as well as significant number of other nanodiamonds \cite{stehlik2015size,yoshikawa1995raman,mermoux2018raman}, is presence of the non-diamond phase giving the strong Raman signal in the wide band around 1580 cm$^{-1}$. The non-diamond carbon is typically ''black'' and strongly  absorbing, and as it will be seen later such absorption is an essential component of the nanodiamond optical properties.

The additional centrifugation-based fractionation for tuning the fraction of the agglomerates was performed as follows. All three powders were dispersed in demineralized (deionized) water by ultrasonic treatment. The initial concentration of nanodiamond in water was ca. 1 wt. \%. After dispergation, the resulting hydrosols were centrifuged at 18000g for 40 minutes (Sigma 3-30KS centrifuge). In each capsule for centrifugation, a hydrosol has a volume of approx. 6 ml. Thus, primary 4 nm crystallites that did not settle during the centrifugation process and larger particles (agglomerates) were separated. The supernatants recovered after centrifugation are referred as DND Z01 (0.08 wt\%), Z+1 (0.44 wt\%) and Z-1 (0.35 wt\%) hydrosols. The precipitates diluted with demineralized water and ultrasonically treated are DND Z02 (0.58\% by weight), Z+2 (1.07\% by weight) and Z-2 (1.28\% by weight). Their concentrations (WT1) were measured by drying 10 g of each hydrosol, followed by measuring the mass of the sediment on an analytical scale SartoGosm CE-124C. Finally, the additional dilution of the hydrosols was done to achieve the abosrbance values of ~0.3, most suitable for optical measurements due to lowering the effects of multiple scattering and reabsorption. The weight fractions after dilution are designated as WT2. The corresponding data is listed in Table \ref{table1}. Size distributions were obtained using the Malvern ZetaSizer device.

\begin{table}[h]
\caption{Weight fractions of nanodiamonds in hydrosols. WT1 - after centrifugation. WT2 - after dilution and before optical measurements. The concentrations of primary particles $n_P$ and agglomerates $n_1$ and $n_2$ obtained from the simulations described below are also given. \label{table1}}

\begin{tabular}{|l|c|c|c|c|r|}
  \hline
  Sample & WT1, \% & WT2, \% & $n_P$, cm$^{-3}$ & $n_1$, cm$^{-3}$ & $n_2$, cm$^{-3}$ \\
  \hline
  Z+1 & 0.44 & 0.029 & $2.5\cdot10^{15}$ & $4.3\cdot10^{9}$ & $3.3\cdot10^{6}$\\
  Z+2 & 1.07 & 0.0048 & $2.1\cdot10^{14}$ & $2.7\cdot10^{10}$ & $1.8\cdot10^{8}$\\
    Z-1 & 0.35 & 0.023 & $2.0\cdot10^{15}$ & $1.7\cdot10^{9}$ & $1.3\cdot10^{6}$\\
  Z-2 & 1.28 & 0.0074 & $4.6\cdot10^{14}$ & $2.3\cdot10^{10}$ & $1.5\cdot10^{8}$\\
    Z01 & 0.08 & 0.019 & $1.6\cdot10^{15}$ & $1.5\cdot10^{9}$ & $1.6\cdot10^{7}$\\
  Z02 & 0.58 & 0.0059 & $3.7\cdot10^{14}$ & $2.1\cdot10^{10}$ & $2.1\cdot10^{8}$\\
  \hline
 
\end{tabular}
\end{table}

\subsection{Measurements of absorbance spectra without IS and with IS}

\label{ssec-uvvis}

The standard measurements of Abs spectra without IS were conducted with the single beam UV-Vis spectrophotometer Unico SQ2800. For measurements with IS, the double beam UV-Vis spectrophotometer Shimadzu-2450 was used (with ISR-3100 IS Attachment).

\begin{figure}
\centering
\includegraphics[width=0.44\textwidth]{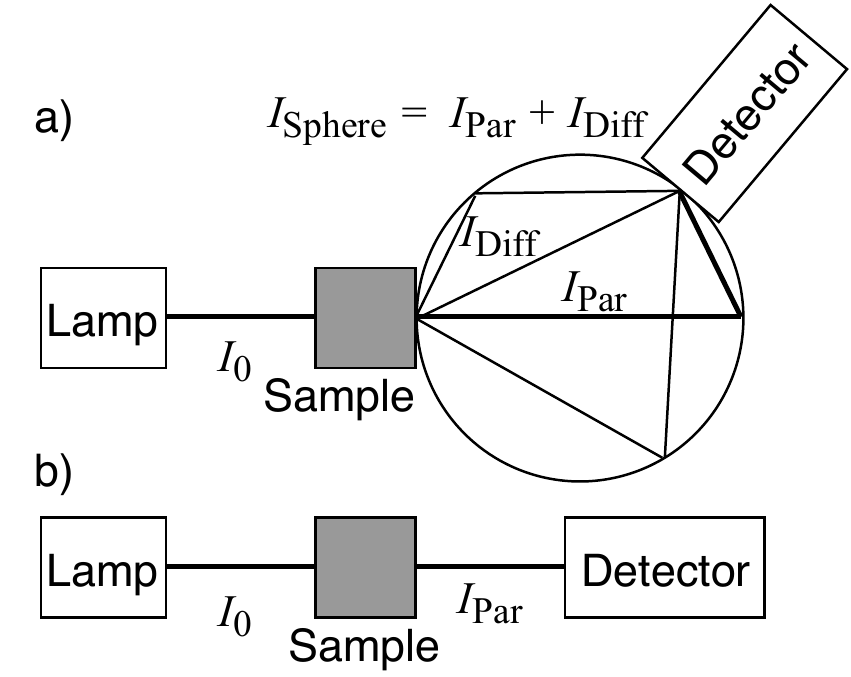}
\caption{\label{fig_sphere} Absorbance spectra measurements on a) spectrophotometer equipped with IS and b) the spectrophotometer without sphere (conventional Abs measurements). For the experiment without sphere, the forward scattered light vanishes, whereas the sphere collects it and brings the additional intensity $I\fs$ into account.}
\label{fig_AbsScheme}
\end{figure}

According to Fig. \ref{fig_AbsScheme}, one can write the following relations for light intensities and values of absorbance without and with sphere:

\begin{eqnarray}
\label{eq_1}
I\WO\ml = I_{\mathrm{0}}\ml \cdot 10^{-\AbsWO\ml},\\
\label{eq_2}
I\s\ml = I\WO\ml + I\fs\ml = I_{\mathrm{0}}\ml \cdot 10^{-\AbsW\ml},
\end{eqnarray}
where $\AbsWO\ml$ and $\AbsW\ml$ are the Abs spectra measured normally (without integrating sphere) and with the integrating sphere, respectively. $I_{\mathrm{0}}\ml$ is the intensity of the incident beam. $I\WO\ml$ is the residue of the incident parallel beam measured in normal experiments without integrating sphere and $I\fs\ml$ is the diffuse light fraction gone to the sphere through the front of the cuvette.


To compare the Abs measurements with the 90 angle scattering experiment described below, let us introduce the light scattering effectiveness as:
\begin{equation}
T\fs\ml=\frac{I\fs\ml}{I_{\mathrm{0}}\ml}
\end{equation}
It is denoted with $T$ because it is defined similar to the transmission coefficient. So, $T\fs\ml$ has the meaning of a light fraction gone out of the cuvette apart the main optical axis and collected by IS. From Eqs. \eqref{eq_1} and \eqref{eq_2} one obtains $T\fs\ml$ as:
\begin{equation}
\label{eq_TscaFromSphere}
T\fs\ml = \left( 10^{-\AbsW \ml} - 10^{-\AbsWO \ml}  \right),
\end{equation}

\subsection{Scattering at 90 degree angle}

\label{ssec_90deg}

\begin{figure}
\centering
\includegraphics[width=0.34\textwidth]{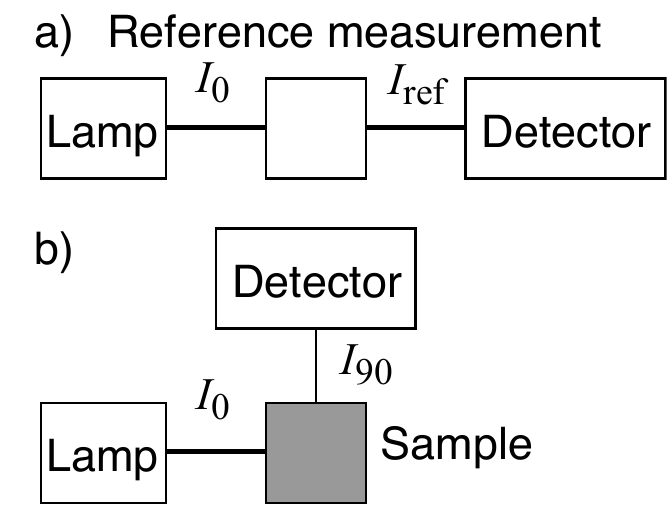}
\caption{\label{fig_chira} 90 degree scattering experiment on Chirascan device giving $T_{90}\ml$. Reference measurement was carried out with water in the cuvette.}
\label{fig_ChiraScheme} 
\end{figure}

The Applied Photophysics Chirascan specrophotometer allows setting the photomultiplier tube detector at the angles of 0 and 90 degrees with respect to the incident light direction (see Fig. \ref{fig_ChiraScheme}). At 0 degrees, the reference intensity was measured. Setting the detector at 90 degrees allowed measurements of the relative light intensity, scattered and gone out of the cuvette through its side wall. The experiment yields the 90 degree scattering in terms of effective transmission
\begin{equation}
\label{eq_T90_Chira}
T_{\mathrm{90}}(\lambda)= \frac{I_{90}\ml}{I_{\mathrm{ref}}(\lambda)}
\end{equation}

Both $T\fs\ml$ and $T_{90}\ml$ are generally proportional to the scattering in the hydrosol. The difference is in the scattering direction (scattering indicatrix should be taken into account) and in the detector solid angle. $T\fs\ml$ and $T_{90}\ml$ can be plotted in the same figure for comparison.

\section{Theory and simulation}

To deeper understand the roles of absorption and scattering and separate  their contributions to the nanodiamond hydrosols Abs spectra at a qualitative level, a simulation of the photon random walk in the cuvette was performed. This simulation allows establishing the interconnection between the scattering and the absorption cross sections, the DND concentrations, the Abs spectra without and with the IS, and the intensity of scattering at 90 degrees.

\label{sec_simulation}
\subsection{Size distribution, dielectric permittivity, and fractal structure of nanodiamonds}

Here, we describe how the cross sections and the scattering indicatrices were obtained for the simulations. We have used the trimodal size distribution for nanoparticles in the hydrosol (three typical sizes of the diamond nanoparticles) lying in agreement with the results of DLS measurements (see Figs. \ref{fig_ZP_Size} and S1).

First, the primary crystallites of the size $D_P = 4$ nm with the particles per volume concentration $n_P$ exist in the hydrosol. Then, we consider the medium agglomerates of the size $D_1=90$ nm and concentration $n_1$. The third type of nanoparticles is the large agglomerates of the size $D_2=700$ nm (concentration $n_2$). The adopted size of the primary nanodiamond crystallites $D_P=4$ nm is reported in numerous papers on the detonation nanodiamonds structural properties \cite{osswald2009phonon,vul2006direct,aleksenskii1997diamond,aleksenskii1999structure,shenderova2011nitrogen,ozerin2008x}. The fraction of 4 nm nanodiamonds is hardly visible in the DLS data because the scattering cross section is proportional to the sixth power of nanoparticle size. Only centrifugation in very hard conditions can make the scattering from single diamond nanoparticles dominant\cite{koniakhin2018ultracentrifugation}.

The dielectric permittivity of the primary crystallites was taken in the form

\begin{equation}
\label{eq_eps_primary}
\varepsilon_P \ml = 5.7 +  (\lambda / \lambda_0)^{-3} + i A_P \cdot (\lambda / \lambda_0)^{P_P},
\end{equation}
where $\lambda_0 = 300$ nm was chosen for natural normalization and $A_{P}=0.17$, and $P_P=-3$ are the adjustment parameters (these values give the best fit of the experimental data). Their values were obtained preeminently by fitting the Abs spectra of Z+1 and other supernatants, see section \ref{sec_discussion} for more details. The first two terms of the equation above  with sufficient accuracy fit the dielectric constant of the bulk diamond, given in Fig. 1 of Ref. \cite{edwards1981infrared}. The value of the diamond dielectric constant is also given in Ref. \cite{bhagavantam1948dielectric}. The third term gives the imaginary part that is essential for absorption. Due to the fact that we interpret the experimental data in the range between 250 nm and 650 nm, Eq. \eqref{eq_eps_primary} should be considered as a local approximation and thus Kramers-Kronig relation \cite{landau2013electrodynamics} is not applicable to it.

The dielectric permittivity of the agglomerates does not match with one of the primary crystallites because they have a sparse fractal-like structure and contain the extensive voids filled with the medium (water). We have used the following mixing rule for calculating the agglomerate dielectric function:

\begin{equation}
\label{eq_eps_aggr1}
\varepsilon_A (D,\lambda) = \varepsilon_{A0} \ml  \cdot F(D) + \varepsilon_{W} \cdot (1-F(D)) 
\end{equation}
where $F(D)$ is the filling factor, $\varepsilon_{W}$ is the dielectric constant of water, and 
\begin{equation}
\label{eq_eps_aggr2}
\varepsilon_{A0} \ml = 5.7 +  (\lambda / \lambda_0)^{-3} + i A_A \cdot (\lambda / \lambda_0)^{P_A},
\end{equation}
where  $A_{A}=0.4$ and $P_A=-1$. The detailed analysis of calculating dielectric functions of mixtures is given in Refs. \cite{reynolds1957formulae,karkkainen2000effective}.

The filling factor can be obtained on the basis of the agglomerate size $D$ and the size of a primary crystallite $D_P$ via the formula\cite{sorensen2001light}:
\begin{equation}
F(D) = C_F (D/D_P)^{3-D_f},
\end{equation}
where $D_f$ is the fractal dimension of agglomerates. The used fractal dimension $D_f=2.45$ coincides with the neutron scattering data on the spatial structure of the DND agglomerates listed in Table 1. in Ref. \cite{tomchuk2014structural}. $C_F=1.9$ was an adjustment parameter. See also Ref. \cite{baidakova1999ultradisperse} for the data on DND fractal structure. While these studies give the fractal dimension only for the agglomerates of the size of approximately 100 nm, the self similarity allows us to  extend these values to the large agglomerates of the employed trimodal model.

The difference in the imaginary part of the dielectric permittivity for the primary crystallites and agglomerates should not be surprising. As discussed below, the absorption (defined by the imaginary part of $\varepsilon$) in the primary particles and in the agglomerates can take place in the carbon phases of various nature. Moreover, Refs. \cite{draine1984optical,gioti2003dielectric,vantarakis2009interfacial,edwards1981infrared} show that the dielectric properties of various carbon allotrope forms differ dramatically and thus some arbitrariness in the choice of $\varepsilon$ is allowed.

The total diamond mass fraction in the hydrosol (WT2 column of Table 1) writes as

\begin{equation}
{\rm WT2} = \frac{1}{8}\frac{4 \pi \rho_D}{3\rho_W} \left(n_P D_P^3 + n_1F(D_1) D_1^3 + n_2F(D_2) D_2^3 \right)
\end{equation}

\subsection{Theory of light extinction in nanodiamond hydrosols}

The Mie theory \cite{mie1908beitrage,hulst1957light,bohren1983absorption} was used to obtain the absorption and scattering cross sections. The calculations performed in the \textsc{Wolfram Mathematica} package \cite{mathematica2009wolfram} code exaclty reproduce the results of the Matzler \textsc{Matlab} code \cite{matzler2002matlab}.

The absorption and scattering cross sections of the primary crystallites are $\sigma_P\absu\ml$ and $\sigma_P\scau\ml$, respectively. The input parameters for the Mie theory were the size $D_P$, nanoparticle dielectric permittivity $\varepsilon_P \ml$, medium dielectric permittivity $\varepsilon_W$ and wavelength $\lambda$. For the medium agglomerates, the input parameters for the Mie theory were the size $D_1=90$ nm, dielectric permittivity $\varepsilon_A(D_1,\lambda)$, mean dielectric permittivity $\varepsilon_W$, and the wavelength $\lambda$. The yield is the absorption and scattering cross sections $\sigma_1\absu\ml$ and $\sigma_1\scau\ml$, respectively. For the large agglomerates, the input parameters were $D_2=700$ nm, $\varepsilon_A(D_2,\lambda)$, $\varepsilon_W$ and $\lambda$, and the yield was $\sigma_2\absu\ml$ and $\sigma_2\scau\ml$. The Mie theory also gives the scattering indicatrix used in the next section. The example of such indicatrices is plotted in Fig. S2.

The attenuation coefficient in the hydrosol due to the scattering can be written as
\begin{equation}
\label{eq_Ascattering}
A\scau\ml=n_1 \sigma_1\scau\ml + n_2 \sigma_2\scau\ml,
\end{equation}
and the attenuation coefficient due to absorption can be written as
\begin{equation}
\label{eq_Aabsorption}
A\absu\ml=n_P \cdot \sigma_P\absu\ml + n_1 \cdot \sigma_1\absu\ml + n_2 \cdot \sigma_2\absu\ml.
\end{equation}

Finally, the conventional absorbance (or total extinction) of the hydrosol can be written using the attenuation coefficients given by Eqs. \eqref{eq_Ascattering} and \eqref{eq_Aabsorption}:

\begin{equation}
\label{eq_AbsWO}
\AbsWO\ml=\frac{A\scau\ml \cdot X+A\absu\ml \cdot X}{\ln10},
\end{equation}
where $X$ is the optical path in the cuvette.

The described model provides the best balance between the amount of free parameter (which should be kept as low as possible) and the quality of the fit of the experimental data. Using 3 different sizes is a minimal model for the description of the optical properties of nanodiamonds. 4 nm primary crystallites are the basic nanodiamond "bricks". Agglomerates of the characteristic size 90 nm are important for relatively isotropic part of scattering, evident from the $T_{90}\ml$ measurements. The presence of the $\approx700$ nm agglomerates leads to the effect of forward scattering and thus they affect the measurements with the integrating sphere.

As an alternative to Eq. \eqref{eq_eps_aggr1} for deriving the agglomerates dielectric permittivity, one can use the Maxwell-Garnet formula, one of the Hashin-Shtrikman bounds or Wiener bounds, see Ref.\cite{karkkainen2000effective}. However, it will not affect significantly  the decomposition of total absorbance into absorption and scattering. The same is valid for varying the sizes $D_{1,2}$, fractal dimension, and $C_F$. The appropriate values of $n_P$, $n_1$, and $n_2$ can be found to reproduce Abs spectra, $T\fs\ml$ and $T_{90}\ml$. 

\subsection{Photon random walk simulation}

It is impossible to interpret the spectra $\Abs\ml$ obtained with the IS using only Eqs. \eqref{eq_Ascattering} and \eqref{eq_Aabsorption} or similar simple equations. To study the light propagation in the cuvette with the hydrosol which strongly scatters and absorbs light, more complicated approaches are required. First, the photon random walk simulations can be performed. The approach based on random walks simulation is also essential for theoretical interpretation of the experiments in terms of $T_{90}\ml$. Second, the theory of light propagation in turbid media can be used. Currently the family of such theories is known as Kubelka-Munk theory \cite{kubelka1931zeit,kubelka1948new,kubelka1954new}. In the straightforward implementation such theory can describe the $T_D$, but not $T_{90}$ and thus to interpret all experimental data, the photon random walk simulations are necessary.

Figures S3 and S4 present the geometry of simulations. During the simulation, the photon starts in the center of the left wall of the cuvette. The propagation direction is along the optical axis (normal to the left cuvette wall). With the probability $n_1 \sigma_1\scau\ml$, the photon is scattered by the medium agglomerates and changes its propagation direction according to the scattering indicatrix calculated with the Mie theory. The same is for large agglomerates (index 2). The indicatrices are given in the supplementary materials \cite{suppl}. Has the photon been scattered or not, it is moved by the small distance $dl=0.3$mm along its actual propagation direction. Also, at each step the photon can be absorbed with a probability $A\absu\ml dl$. In this case the simulation stops and goes to the next photon. The total amount of photons simulated for each wavelength was $N_{\rm total}=10^5$. If the photon reaches the cuvette wall, the simulation stops and also goes to the next photon.

Some areas at the walls correspond to detectors. So at the side wall (parallel to the optical axis) there was a ''Chirascan detector'' area of the size 0.64 cm. Taking an amount of photons fallen onto the ''Chirascan detector'' areas $N_{90}$, one can write $T_{90}=\alpha N_{90}/N_{\rm total}$, where $\alpha$ was an adjustment parameter related with the actual solid angle of the detector.

The simulation allows obtaining  $\Abs_S\ml$  as $-\log_{10}(N\s/N_{\rm total})$, where $N\s$ is amount of photons fallen at the right side (the side opposite to the entrance point). The conventional absorbance was obtained as $-\log_{10}(N_{\rm forw}/N_{\rm total})$, where $N_{\rm forw}$ is the  amount of photons fallen onto a ''normal detector'' area, opposite to the entrance point. The forward scattering efficiency is $T\fs\ml = (N\s - N_{\rm forw})/N_{\rm total}$. 

It is important to note that the Abs spectra can be either calculated analytically using Eq. \eqref{eq_AbsWO} or obtained with photon random walk simulation. The results of analytical calculations and photon random walk simulations coincide with sufficient accuracy (see the supplementary \cite{suppl}, Fig S5). In Figs. \ref{fig_AbsZP_1}, S6, and S9, we plot the Abs spectra using the random walk simulations.

\subsection{Theory of light propagation in turbid media}

Here we use the theory of light propagation in turbid media given by Ryde in Refs.  \cite{ryde1931scattering,ryde1931scattering2}. It was originally written for the single type of nanoparticles, and here we extended it to the case of trimodal particle size distribution. To keep the clarity, we keep the designations from Ref. \cite{ryde1931scattering} and than express them via the ones used in this paper.

The light transmitted through the cuvette, including both the diffuse $T\fs$ light fraction and the residue $T_{\rm Par}$ of the incident parallel beam writes as

 \begin{multline}
    T_{\rm Par} + T\fs = \frac{QK+P \exp(-q'X)NB\sinh(KX)}{(\mu + NB)\sinh(KX)+K\cosh(KX)} \\ - (Q-1)\exp(-q'X),
\end{multline}
where $q'$ is the total attenuation coefficient and other designations are
\begin{equation}
\label{eq_K}
    K = \sqrt{\mu(\mu+2NB)},
\end{equation}
\begin{equation}
\label{eq_Q}
    Q = \frac{2\mu F'+N(B+F')(F'+B')}{2\mu F + N(F'+B')^2},
\end{equation}
\begin{equation}
\label{eq_P}
    P = \frac{N(B-B')(F'+B')}{2\mu F + N(F'+B')^2},
\end{equation}
In the equations above $\mu$ is the absorption coefficient of the medium: $\mu \equiv A\absu\ml$ and $N$ is the concentration of the particles. The quantities $F'$ and $B'$ are the forward and backward scattering cross sections. Their definition assumes the collimated illumination of the nanoparticle and integrating the light scattered to the forward and backward semi-spheres, respectively. The quantities $B$ and $F$ are defined as follows. $B$ is the total quantity of light which is scattered to the backward semi-sphere when the particle is illuminated equally from all directions from the backward semi-sphere. $F$ is the total quantity of light which is scattered to the forward semi-sphere when the particle is illuminated equally from all directions from the backward semi-sphere. Here, $F'$, $B'$, $B$, and $F$ were calculated based on the Monte Carlo integration of the scattering indicatrices obtained with the Mie theory. By the definition $B + F = B' + F' = \sigma\scau$. In the Rayleigh limit (particle size is much smaller than the wavelength) the scattering intensity is symmetric with respect to the plane perpendicular to the incident beam and one writes $B=F=B'=F'$. In the case of nanoparticles comparable by size with the wavelength (Mie limit), one has strong dominating of forward scattering $F'\gg B'$.

As mentioned above, this approach can be naturally used for the hydrosol with several types of particles. So, for the designations used in our paper, to derive e.g. $F'$ one should perform the following replacement in Eqs. \eqref{eq_K}, \eqref{eq_Q} and \eqref{eq_P}:
\begin{equation}
    NF' \rightarrow n_1 \cdot F'_1 + n_2 \cdot F'_2,
\end{equation}
where $F'_1$ and $F'_2$ are the forward scattering cross sections for medium and large agglomerates, respectively. The same substitutions are to be done for $B'$, $B$, and $F$.





\section{Results}

\label{sec_results}

\subsection{Experimental results}
The results of the size measurements for supernatants and precipitates of all samples obtained with DLS are shown in Figs. \ref{fig_ZP_Size} and Fig. S1 in the supplementary data\cite{suppl}. Fig. \ref{fig_expAbs} shows the Abs spectra of all samples measured without and with IS. Fig. \ref{fig_expTsca} shows the scattering efficiency in terms of $T\fs\ml$ obtained from the Abs spectra measurements using Eq. \eqref{eq_TscaFromSphere} and $T_{90}\ml$ from the 90 degree scattering experiment on Chirascan.

\begin{figure}[h]
\centering
\includegraphics[width=0.5\textwidth]{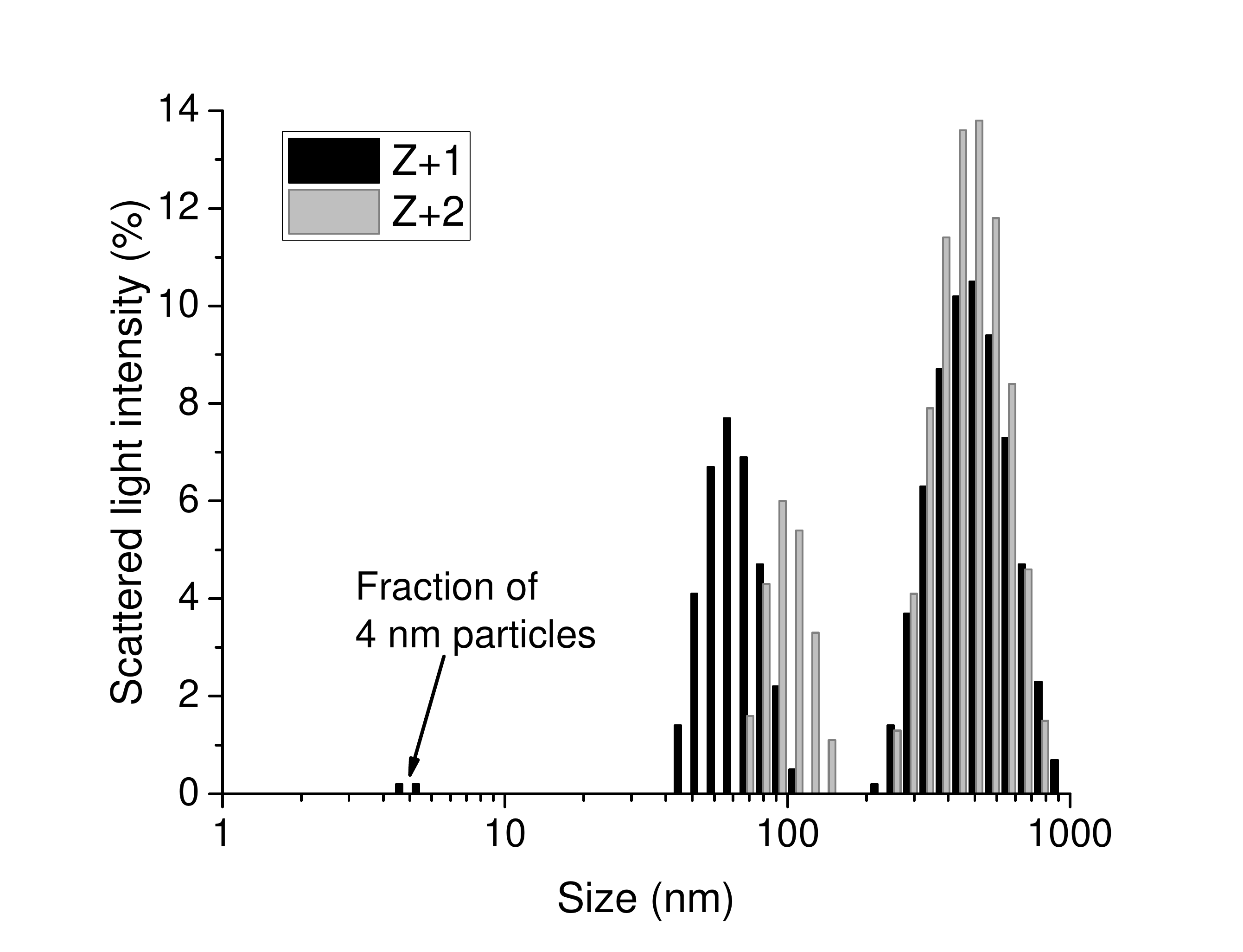}
\caption{\label{fig_ZP_Size} Distribution by scattered light intensity for Z+ nanodiamond supernatant (Z+1) and precipitate (Z+2) obtained by DLS. One sees the trimodal size distribution.}
\end{figure}

One sees that for supernatant deagglomerated diamond types (DND Z+1, Z-1), the Abs spectra without and with IS nearly coincide. It means that the scattered light intensity $I\fs$ is small and the main contribution arises from the absorption. On the contrary, for all precipitates (samples Z+2, Z-2, Z02) the difference with and without the sphere is significant. The difference is also tangible for the Z01 sample, because it lacks the deagglomeration procedure and intensively scattering agglomerates remain in the hydrosol. Thus, the centrifugation process leads to separation and manifestation (due to absorption) of fraction smaller than 100 nm.

The spectra of $T\fs\ml$ ans $T_{90}\ml$ correlate with the Abs spectra without and with the sphere. Again, the scattering (both forward $T\fs\ml$ and at 90 degree angle $T_{90}\ml$) from the supernatants (the samples with the index 1) is very low and the scattering from the precipitates (the samples with the index 2) is at least one order higher. Z01 exhibits an intermediate case.

\begin{figure}[h]
\centering
\includegraphics[width=0.5\textwidth]{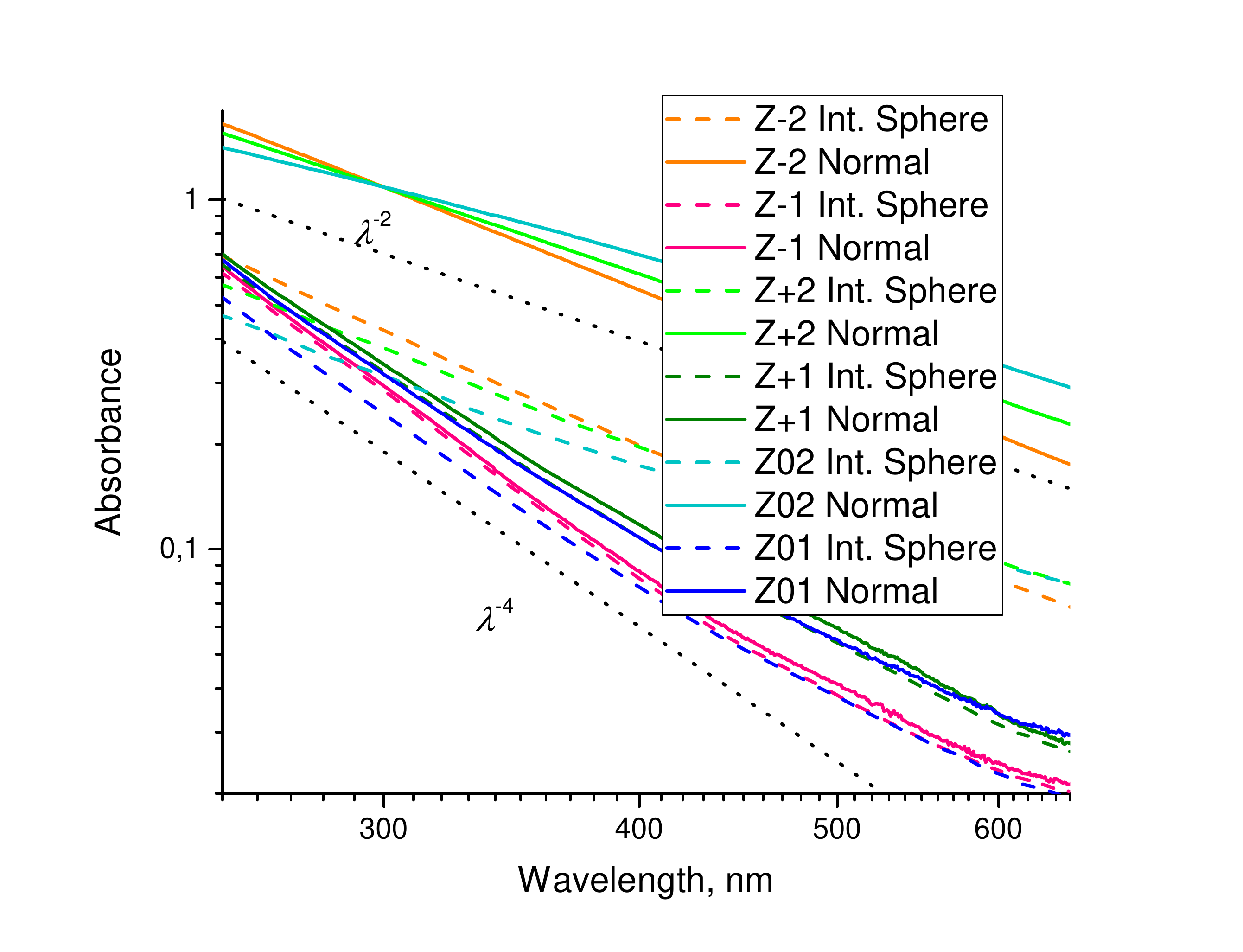}
\caption{\label{fig_expAbs} Absorbance spectra for all the studied samples. Orange is for Z-2, pink is for Z-1 sample, light green is for Z+2, dark green is for Z+1, cyan is for Z02, and blue is for Z01. Solid curves are for $\AbsWO\ml$ (measurements without IS) and dashed curves are for and $\AbsW\ml$ (measurements with IS). Also, the  $\lambda^{-2}$ and $\lambda^{-4}$ functions are plotted with dotted black curves.} 
\end{figure}

\begin{figure}[h]
\centering
\includegraphics[width=0.5\textwidth]{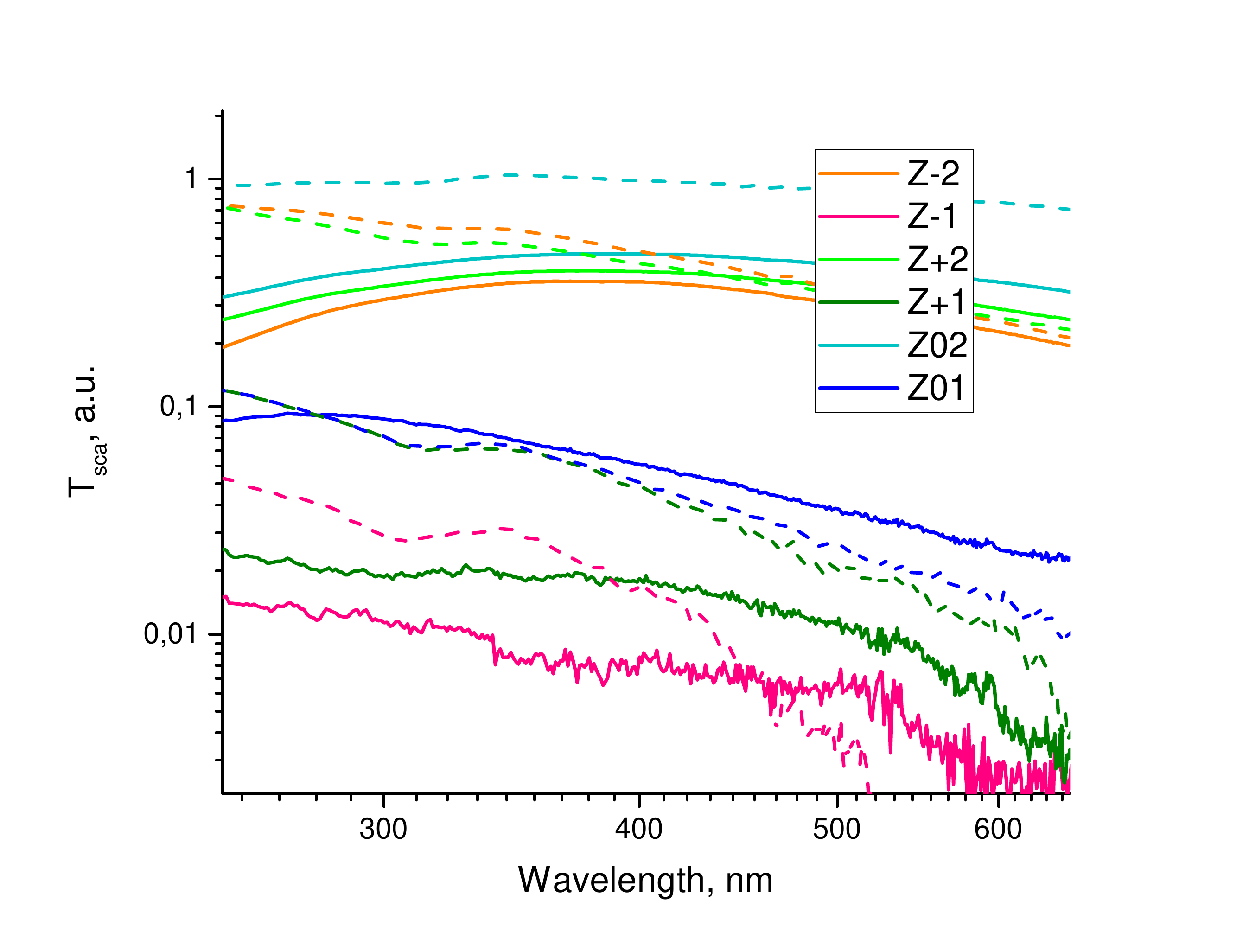}
\caption{\label{fig_expTsca} Scattering effectiveness for all samples in terms of $T\fs\ml$ and $T_{90}\ml$. Solid curves are for $T\fs\ml$ obtained from Abs measurements without and with sphere substituted to Eq. \eqref{eq_TscaFromSphere}. Dashed curves are for 90 scattering experiment on the Chirascan device, namely $T_{90}\ml$. Orange color is for Z-2 sample, pink is for Z-1 sample, light green is for Z+2, dark green is for Z+1, cyan is for Z02, and blue is for Z01.}
\end{figure}

Above 600 nm, the signal (and especially scattering) from the supernatants becomes too weak and comparable with the device sensitivity for both used VU-vis spectrophotometers and Chirascan.

The described analysis of Abs spectra without and with IS and the scattering efficiency ($T\fs\ml$ and $T_{90}\ml$) provides the possibility to  estimate the contributions of absorption and scattering to the light extinction in nanodiamond hydrosols only qualitatively. The same data accompanied with the theory and random walk simulations allow more precise quantitative approach for the separation of absorption and scattering contributions. Additional information can be obtained  on the nanoparticle size, the agglomerates fraction, and the dielectric properties of primary crystallites and agglomerates.


\subsection{Comparison of experimental data with the results of photon random walk simulation}

\begin{figure*}
\includegraphics[width=0.7\textwidth]{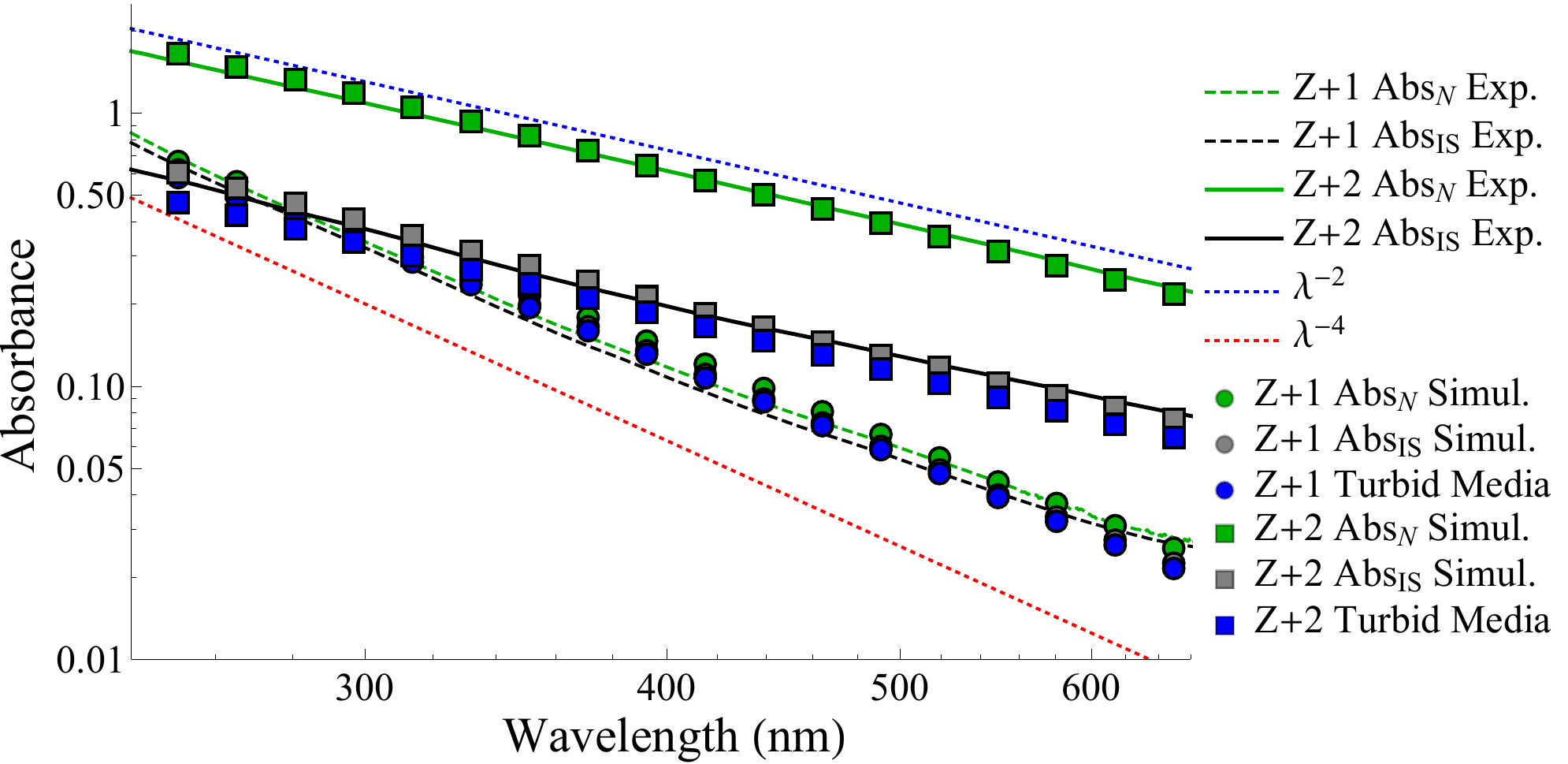}
\caption{\label{fig_AbsZP_1} The Abs spectra with and without sphere, obtained with photon random walk simulation (depicted with markers) and the experimentally measured Abs spectra of Z+1 (dashed curves) and Z+2 (solid curves) samples with (black curves) and without the integrating sphere (green curves). The predictions of the theory of light propagation in turbid media for Abs spectra with integrating sphere are shown with the blue markers. The red dashed curve shows the $\lambda^{-4}$ function corresponding to Rayleigh scattering and the blue dashed curve is for $\lambda^{-2}$.}
\end{figure*}

\begin{figure*}
\centering
\includegraphics[width=0.7\textwidth]{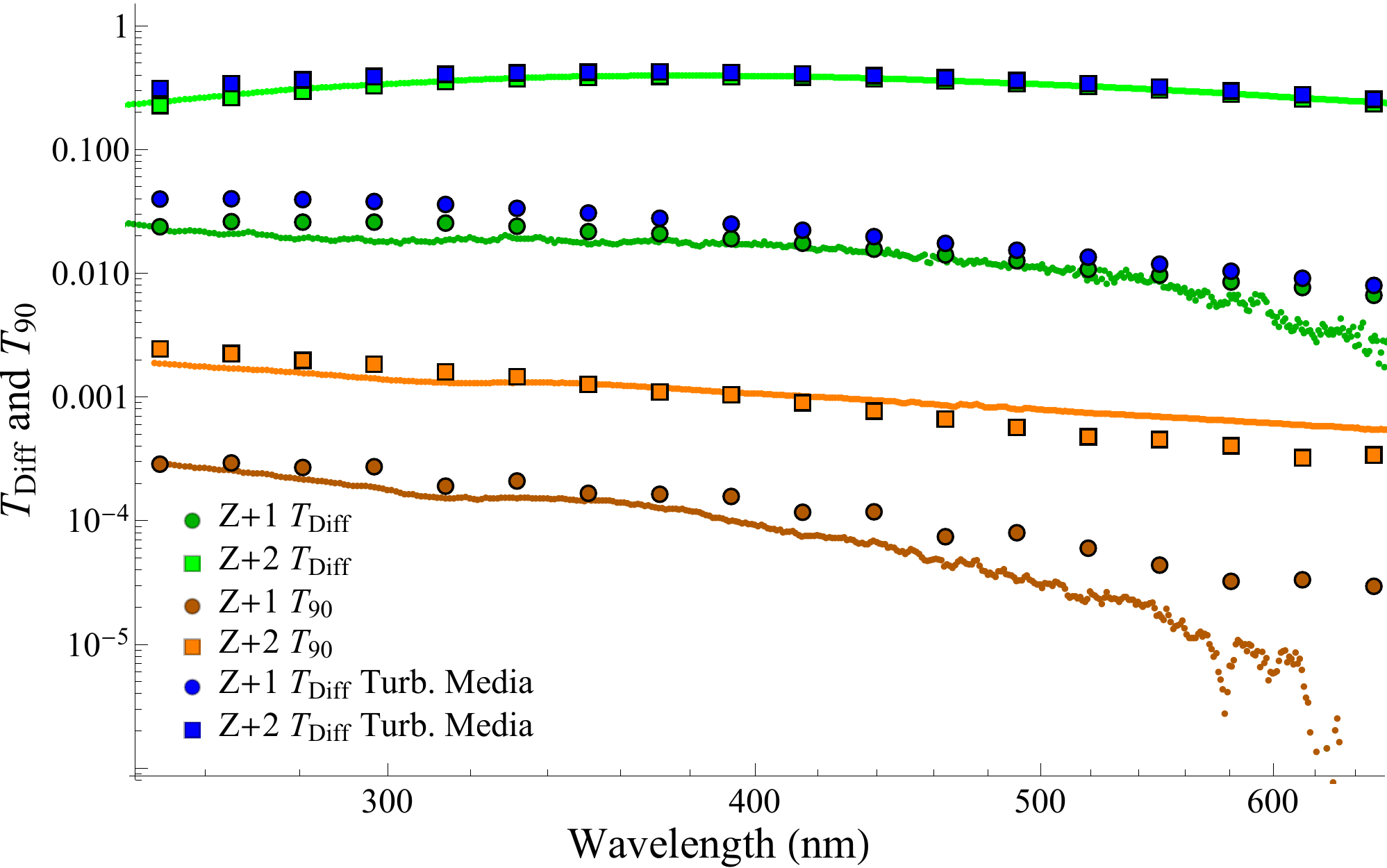}
\caption{\label{fig_TscaZP} The results of photon random walk simulations for scattering efficiency in terms of $T\fs\ml$ (green colors) and $T_{90}\ml$ (orange colors) are shown with markers. Circles are for Z+1 and squares are for Z+2 sample. The experimental spectra are given by dense points. Experimental $T\fs\ml$ spectra were obtained on the basis of Abs measurements with and without sphere using Eq. \eqref{eq_TscaFromSphere} and experimental $T_{90}\ml$ spectra were obtained from 90 degree scattering experiment on Chirascan device using Eq. \eqref{eq_T90_Chira}. Blue markers denote the results of the theory of light propagation in turbid media.}
\end{figure*}

Fig. \ref{fig_AbsZP_1} shows the  $\AbsWO\ml$ and $\Abs\ml$ spectra of Z+1 and Z+2 samples, calculated on the basis of photon random walks simulations with the best set of adjustment parameters compared to experimental data (see Fig. S6 and S9 for Z- ans Z0 samples, respectively). The results obtained by the theory of light propagation in turbid media are also given. Fig. \ref{fig_TscaZP} shows the scattering efficiency in terms of $T\fs\ml$ and $T_{90}\ml$  (see Fig. S7 and S10 for Z- ans Z0 samples, respectively). The parameters $D_f$ (fractal dimension), $A_P$, $P_P$, $A_A$, and $P_A$ (constants in dielectric permittivity), $C_F$ as well as the sizes $D_P$, $D_1$, and $D_2$ were the same for all samples (DND Z+1, Z+2, Z-1, Z-2, Z01, and Z02). For each sample $n_P$, $n_1$, and $n_2$ were adjusted separately.

\begin{figure*}
\includegraphics[width=0.9\textwidth]{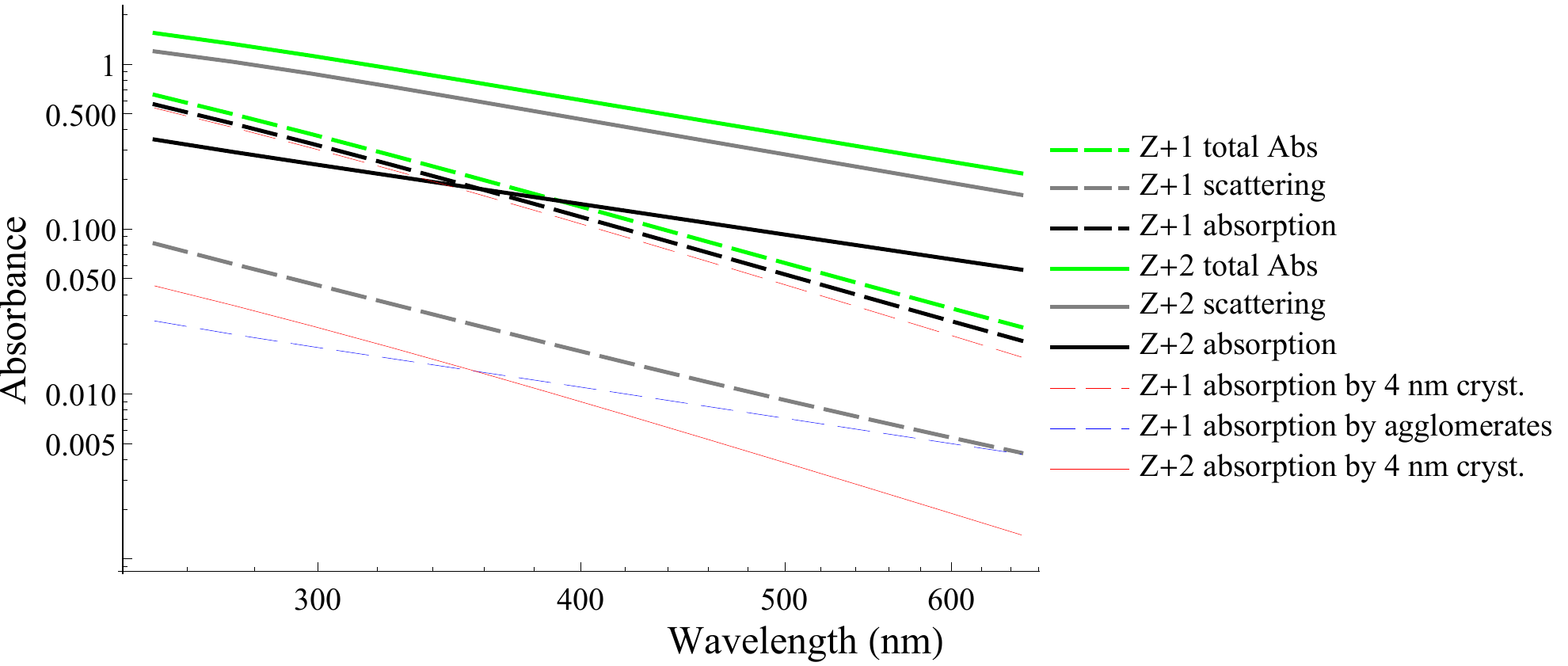}
\caption{\label{fig_Decomp} Decomposition of the Abs spectra to scattering and absorption based on the Eqs \eqref{eq_Ascattering} and \eqref{eq_Aabsorption}. Solid curves are for Z+2 and dashed curves are for Z+1 samples. Green color is for total Abs, gray color is for scattering and black color is for absorption. For supernatant Z+1, the contribution to absorption from primary crystallites (red color) and agglomerates (blue color) is given. For precipitate Z+2, both scattering and absorption is governed mainly by agglomerates.}
\end{figure*}

Fig. \ref{fig_Decomp} is the main result of present paper. It shows the Abs spectra decomposition into scattering and absorption contributions. Namely, Fig. \ref{fig_Decomp} shows the Abs spectra of absorption and scattering obtained using Eqs. \eqref{eq_Aabsorption} and \eqref{eq_Ascattering}, respectively for Z+1 and Z+2 samples. The concentrations were  adjusted and the cross sections were obtained by the Mie approach as described below. The figures plotting the similar decomposition for Z- and Z0 samples are given in supplementary (Figs. S8 and S11, respectively).

\section{Discussion}
\label{sec_discussion}
\subsection{Quantifying absorption and scattering contributions}

One sees a good agreement between the predictions of the the photon random walk simulations, theory of light propagation in turbid media, and the experimental results. For the supernatant Z+1, the scattering is one order smaller than the absorption. The scattering in governed by the agglomerates remained in the hydrosol only. The absorption is predominantly due to primary crystallites (it is approximately 3 times higher than the absorption from the agglomerates). For the precipitate Z+2, the scattering dominates and it is several times larger than the absorption. The scattering is obviously governed by the agglomerates. The absorption is also due to agglomerates (typically it is 10 times higher than the absorption from the primary crystallites). Thus, the optical properties of precipitates are completely defined by the agglomerates.

Interestingly, the absorption in the Z+1 sample and other supernatants is accidentally closer by its slope to the Rayleigh scattering (giving the famous $\lambda^{-4}$ for the scattering cross section) than the true scattering in Z+2 sample and other precipitates in the Mie limit. That is the reason why previously\cite{vul2011absorption,aleksenskii2012optical,konyakhin2013labeling} the nanodiamonds Abs spectra were treated as follows: first, $\lambda^{-4}$ was subtracted from the spectra as some presumable scattering background, then the remaining signal was attributed to the absorption on amorphous or sp$^2$-like phase. From the present results, one sees that this algorithm is not correct for both precipitates and supernatants (and also for the suspension before centrifugation). Even for precipitates, the scattering contribution never overcomes 90\%. Previously\cite{vul2011absorption}, the agglomerates were considered as solid objects, whereas in the present approach we account for their fractal sparse structure. Nevertheless, the conclusion that the scattering in DND hydrosols is due to the agglomerates (and not due to 4 nm fraction) given in previous works \cite{vul2011absorption,aleksenskii2012optical,konyakhin2013labeling} stays intact. However, one sees that the scattering contribution to OD spectra is much smaller than thought previously.

A very similar picture takes place for the Z-1 and Z-2 samples. For Z01 sample (supernatant) the scattering is one order higher than for Z+1 and Z-1 samples (but it is still several times smaller than absorption). This picture agrees with the fact that Z0 diamond is an initial specie for Z+ and Z- preparation by means of annealing and chemical deagglomeration. According to Table \ref{table1}, the Z01 sample contains larger fraction of agglomerates than Z+1 and Z-1. Z02 sample is a precipitate of non deagglomerated diamond and it should contain a lot of large agglomerates. Thus, the  trimodal model with the fixed sizes, suitable for all other samples, works worse for the Z02 sample. One can conclude that centrifugation is indeed a very effective way to control the optical properties of nanodiamonds \cite{koniakhin2018ultracentrifugation,trofimuk2018effective,usoltseva2018absorption}.

From Fig. \ref{fig_AbsZP_1} one sees that the slopes of the scattering and the absorption for Z+2 sample are the same. It can be explained by the transition to the geometric optics limit caught by Mie theory. In this limit, both scattering and absorption cross sections do not depend on wavelength and they are proportional to the surface of the geometric shadow $\frac{1}{4}\pi D^2$. Importantly, the light wavelength is effectively decreased by the high value of the water refraction index, which helps to approach the limit of geometric optics. For the higher values of agglomerate size $D_2$ and higher fractal dimension $D_f$, one reaches completely the geometrical optics limit with no wavelength dependence in OD spectra (flat spectra), and an agreement of simulation and experiment can not be achieved.

From the results described above, one unambiguously concludes that the accounting for scattering in nanodiamond hydrosols absolutely requires the use of the Mie theory, because it is due to the agglomerates of the size of hundreds nanometers remaining in the solution. Due to the specific interplay between the wavelength, particle size, and dielectric properties of agglomerates (possessing the fractal structure with extensive voids), one observes the rather weird scattering slope indistinguishable from $\lambda^{-2}$ for Z+2 samples. For Z-2 and Z02 the slope is slightly different from $\lambda^{-2}$, see Figs. S6 and S9. The Rayleigh approximation is clearly not enough for the description of such structures. For 4 nm fraction, the  scattering is vanishing with respect to the absorption and can be neglected. This is one of the main outcomes of the present study. The fact that the absorption always dominates or at least gives a significant contribution (dozens of percents) to absorbance, allows measuring the nanodiamond weight concentration directly, as a quantity straightly proportional to the absorbance in UV-vis range (except in the case of large agglomerates presenting specifically in precipitates).

Due to the strong absorption as well as to multimodal and broadened size distribution, the studied hydrosols were a complicated case for investigation by the DLS technique, see e.g. Ref. \cite{aleksenskii2012applicability}. Nevertheless, the Abs spectra measurements ascertained the predictions of DLS. The given by DLS trimodal size distribution was indeed the minimal model to describe the Abs spectra of the studied hydrosols. The large $D_2$ agglomerates have the strongest forward scattering while the medium agglomerates (with the characteristic size of $D_1=90$ nm) have more isotropic scattering indicatrix. It means that the experimental data from the IS are mostly affected by larger agglomerates and for the 90 degree scattering $T_{90}$ measured at Chirascan device, the contribution of medium agglomerates is more important.

Additional information about absoprtion and scattering in DND hydrosols can be obtained e.g. using the angle resolved scattering. However, even the single wave length MADLS (multi-angle dynamic light scattering) devices are not so widespread as common DLS devices. The same picture is actual for such advanced methods as laser calorimetry, photoacoustic or photothermal spectroscopy. The very perspective will be the simultaneous usage of these methods and Abs measurements with integrating sphere to directly compare the main results. In present paper we make an effort to get as much as possible information related to the DND optical properties using the easily accessible and common equipment and thus we restrict ourselves to the usage of integrating sphere. The essential role in our approach is played by the consequential processing of the results using Mie theory followed by either photon random walks simulations or Kubelka-Munk theory. The employed 90 degree scattering configuration is not so common and it was used to supply the main conclusions coming from the analysis of the Abs spectra obtained with the integrating sphere. We show that the usage of the described above mixture of experimental and theoretical approaches is sufficient to fully solve the addressed problem, namely quantifying the scattering and absorption in nanodiamond hydrosols. Finally, photoacoustic method (see e.g. Fig. 2 from Ref. \cite{usoltseva2018absorption}) shows the close by magnitude contribution to Abs spectra from light absorption.

\subsection{Structural properties of nanodiamonds and role of functional groups}

In addition to justification the fact that the absorption is a dominant light extinction mechanism in the nanodiamond hydrosols, present results also allow some general conclusions about the structure and the dielectric properties of nanodimonds and their agglomerates. First, fitting the experimental data requires the assumption that the agglomerates are not solid and that they contain extensive voids. The fractal dimension 2.4 agrees both with recent SANS data \cite{tomchuk2014structural} and with obtained experimental data.

It is known that absorption bands in the UV area in nanocarbon structures can arise from the presence of oxygen-containing moieties. For instance, absorption feature at 300 nm in graphene oxide is commonly attributed to n-$\pi$* transitions in C=O bonds of carbonyl and caboxyl groups \cite{kumar2014scalable}. However, comparison of the DND Z+ and DND Z- absorption spectra demonstrates that functionalization parameters do not affect the absorption in nanodiamonds. DND Z+ particles are covered mostly with carbohydrate (CH2, CH3) moities with little content of hydroxyls (-OH). On the other hand, DND Z- if predominantly fucntionalized by carboxyls (-COOH) and aldehydes (-COH). The detailed results on the surface chemistry of the studied samples can be found in the Ref.\cite{dideikin2017rehybridization}. Despite such a strong difference in the functionalization parameters, DND Z+ and DND Z- exhibit almost equal absorbance spectra both as individual particles and as aggregates (see Fig. \ref{fig_expAbs}). Furthermore, the optical properties of all samples (Z+,Z-,Z0) can be theoretically reproduced on the basis of the same dielectric permittivity for primary particles P and agglomerates A. Based on these facts, one can formulate the hypothesis that absorption in detonation nanodiamonds is an intrinsic property of nanoparticle lattice (diamond core or reconstructed surface) and supervenient electronic structure.

The agreement between experiment and theory can be achieved only if the imaginary parts of the dielectric permittivities of the agglomerates and the primary crystallites do not coincide: $A_P \neq A_A$ and $P_A \neq P_P$. This fact supports the hypothesis that the absorption takes place in the carbon phases of different nature in primary crystallites and in agglomerates. More specifically, Ref. \cite{dideikin2017rehybridization} shows that sp$^2$ phase forms the linkages between primary DND crystallites in the agglomerates and the deagglomeration is due to removing these linkages. Thus, one can conclude that the sp$^2$ phase can give a significant contribution to absorption in agglomerates. From the value of $A_A$ and the typical magnitude of the black carbon dielectric permittivity imaginary part $\left< \varepsilon'' \right> \approx 9$ (see Fig. 2 in Ref. \cite{draine1984optical}) one can estimate the fraction $f_A$ of the non-diamond phase in the agglomerates as $f_A=\frac{A_A}{\left< \varepsilon'' \right>} \approx 0.05$.

For the primary crystallites, the absorption potentially could arise from the Urbach tail in the electron density of states (due to the disorder) near the band gap edge. However, this hypothesis implies the exponential wavelength dependence of $\varepsilon_P$ imaginary part \cite{vantarakis2009interfacial}. But we did not manage to fit $\mathrm{Im}\{\varepsilon_P\}$ in the exponential form for explaining absorption in the DND Z+1, Z-1, and Z01 samples. The power function with the best fit quality corresponds to $P_P=-3$. The second hypothesis explains the absorption by the non-diamond phase (sp$^{3-x}$ or even graphite-like) shell evidenced by UV-vis and Raman spectroscopy \cite{tomita2002optical,korepanov2017carbon,mermoux2018raman} and by the means of X-ray diffraction and electron diffraction \cite{kulakova2010structure,hawelek2008structural,yur2010x}. Fig. 2 from Ref. \cite{gioti2003dielectric} shows the dielectric permittivities of various types of amorphous carbon differ dramatically, which allows certain arbitrariness when tuning the dielectric permittivity. E.g. approximating with the power function the dependence for ta-C in  Ref. \cite{gioti2003dielectric} one obtains $P_P \approx -2$. Using the adjusted value of $A_P$ one can estimate the effective fraction of graphite-like phase in the primary crystallites as $f_P = \frac{A_P}{\left< \varepsilon'' \right>} \approx 0.02$. Noteworthy, the latter quantity is an essential input parameter for modelling the disorder effects and line width in the nanodiamonds Raman spectra using microscopic DMM-BPM \cite{koniakhin2018raman} or continuous EKFG \cite{utesov2018raman} models. The diffraction studies \cite{kulakova2010structure,hawelek2008structural,yur2010x} indicate thickness of such disordered shell up to 1 nm, which gives drastic volume fraction for 4-5 nm particles. However the shell phase can not be considered as purely sp$^2$ and thus contains smaller effective fraction of sp$^2$ carbon. The extraction of non-diamond phase fraction from Raman measurements does not provide absolute accuracy due to differing scattering cross sections for diamond and graphite components.

\section{Conclusion}

As a net result, it is demonstrated that the preeminent part of the individual DND particles optical spectra is governed by the absorption of light, and not by its scattering. The scattering begins to dominate only for the DND agglomerates with the lateral size of several hundreds of nanometers. Although the exact mechanism underlying the absorption process remains unclear, the obtained results give a deeper understanding of the DND optical properties and allow to clarify the calculations involved in the analysis of the DND fluorescence spectra and particle size using dynamic light scattering. Normal Abs measurements supported by the measurements with IS or by measurements of side scattering allow distinguishing the contributions of scattering and absorption to nanodiamond spectra and can provide a deeper insight into the properties of their surface and phase composition. Clarification of the absorption mechanism in the primary DND crystallites and creation of an appropriate theoretical model is a bright challenge for the future studies of nanodiamonds.

\begin{acknowledgments}

Theoretical and computational contributions, a part of optical experiments as well as developing the general idea of the study were conducted by S.V.K. and funded by RFBR according to the research project 18-32-00069. We acknowledge the project "Quantum Fluids of Light" (ANR-16-CE30-0021). A.V.S. acknowledges RFBR (project 18-29-19125 MK) for sample preparation and DLS measurements. Thanks to A.Ya. Vul for his support. We are gratefully indebted to O. Bleu and D.D. Solnyshkov for useful criticism.
\end{acknowledgments}

\bibliography{dnd}

\end{document}